\documentclass[12pt]{article}
\title{ New developments in nonperturbative QCD}
\author{  Yu.A.Simonov\\
 State Research
Center\\Institute of Theoretical and Experimental Physics, \\
Moscow, 117218 Russia}
\newcommand{\beq}{\begin{eqnarray}}
 \newcommand{\eeq}{\end{eqnarray}}
\newcommand{\be}{\begin{equation}}
 \newcommand{\ee}{\end{equation}}

\def\fun#1#2{\lower3.6pt\vbox{\baselineskip0pt\lineskip.9pt
\ialign{$\mathsurround=0pt#1\hfil ##\hfil$\crcr#2\crcr\sim\crcr}}}

\newcommand{{\SD}}{\rm SD}

\newcommand{\ver}{\mbox{\boldmath${\rm r}$}}

\newcommand{\vep}{\mbox{\boldmath${\rm p}$}}

\newcommand{\lan}{\langle}
\newcommand{\ran}{\rangle}
\begin{document}
\maketitle

\begin{abstract}
Basic developments in the analytic study of the QCD vacuum
structure and of the QCD spectrum, including glueballs and hybrids
are reviewed.
\end{abstract}

\section{Introduction}

One of the main problems of QCD is the development of quantitative
analytic nonperturbative methods. Recently a very promising new
method has been suggested \cite{1,2} (for a review see \cite{3}),
called the Field Correlator Method (FCM) which is fast progressing
now.

FCM has a wide scope and allows to study all nonperturbative
problems  from  the QCD vacuum structure \cite{4,5} to the details
of the QCD spectrum \cite{6}-\cite{9}, and high-energy scattering
\cite{10}. A practical usefulness of the method is connected to
the number of gauge-invariant field correlators which should be
included as the input of the method. Recently the lattice study
\cite{11} has confirmed the phenomenon of the Casimir scaling for
the static $Q\bar Q$ interaction, \cite{5}, which implies  the
domimnant role of the quadratic (Gaussian)  field correlators with
accuracy of order of 1\%. In this way a very simple picture of the
QCD vacuum emerges, called the Gaussian Stochastic Approximation
(GSA), which implies short range correlations between field
strength operators in the vacuum and  the simple spectrum
described by  the only scale parameter -- string  tension $\sigma$
(with the $\alpha_s$ and correlation length $T_g$ defining some
10\% corrections to the spectrum).

In this talk we shall review the situation with the QCD vacuum
structure and  the derivation of the QCD  spectrum.

\section{ The QCD vacuum structure. Stochastic $vs$ coherent.}

The basic quantity which defines the vacuum structure in QCD is
the field correlator (FC)
\be
D^{(n)}{(x_1,...x_n)} \equiv \lan F_{\mu_1\nu_1} (x_1)
\Phi(x_1,x_2) F_{\mu_2\nu_2}(x_2) \Phi(x_2, x_3) ...
F_{\mu_n\nu_n} (x_n) \Phi(x_n, x_1)\ran,\label{1}\ee $$
\Phi(x,y)=P\exp ig \int^x_y A_\mu(z) dz_\mu.$$ The set of FC
(\ref{1}) for $n=2,3,...$  gives a detailed characteristic of
vacuum structure, including field condensates (for coinciding
$x_1=x_2=...x_n$). On general grounds one can distinguish two
opposite situations: 1) stochastic vacuum 2) coherent vacuum. In
the first case FC form a hierarchy with the  dominant lowest term
$D^{(2)} (x_1,x_2) =D^{(2)} (x_1-x_2)$, while higher FC are fast
decreasing with $n$. In the second case  all FC are comparable,
and expansion of physical amplitudes as the series of FC is
impractical. This is the case for the  gas/liquid of classical
solutions, e.g. of instantons, magnetic monopoles etc. The
physical picture behind the situation of nonconverging FC series
is that of  the  coherent lump(s), when all points in the lump are
strongly correlated.

To understand where belongs the QCD vacuum one can start with the
Wilson loop in the representation $D$ of the color group  SU(3),
\be
W_D(C) = \lan tr_D\exp (ig \int_C  dz_\mu  A^a_\mu
T_a^{(D)})\ran\label{2} \ee

The Stokes theorem and the cluster expansion identity allow  to
obtain the basic equation, which is used  in most applications of
the FCM (for more details see \cite{3})
\be
W(C)= \exp \sum_n \frac{(ig)^n}{n!} \int \tilde
D^{(n)}(x_1,...x_n) d\sigma_{\mu_1\nu_1}(x_1)...
d\sigma_{\mu_n\nu_n} (x_n).\label{3}\ee

Here integration is performed over the minimal surface $S_{min}$
inside the contour $C$ defined in (\ref{2}) and $\tilde D_n$ is
the so-called cumulant or the connected correlator, obtained from
the FC Eq.(\ref{1}) by subtracting all disconnected averages. From
(\ref{3}) one easily obtains that the Wilson loop has the area-law
asymptotics, $W(C)\sim \exp (-\sigma S_{min})$, for  any finite
number of terms $n_{max}; n\leq n_{max}$ in  the exponent
(\ref{3}).

The string tension is expressed through $\tilde D^{(n)}$. \be
 \sigma
=\frac12 \int D^{(2)} (x_1-x_2) d^2 (x_1-x_2) +0(\tilde
D^{(n)},n\geq 4 ) = \sigma_2+\sigma_4 +...\label{4}\ee

Eq.(\ref{4}) has several  consequences: 1) confinement appears
naturally for $n=2$, i.e. in the GSA 2) the lack of confinement
can be due to vanishing of all FC, or  due to the special
cancellation between the cumulants, as it happens for the
instanton vacuum \cite{4}, 3) for static quarks in the
representation $D$ of  the color   group SU(3), the string tension
$\sigma^{(D)}_2$ is proportional to the quadratic Casimir factor (
the  Casimir scaling)
\be
\sigma_2^{(D)} = \frac{C_D^{(2)}}{C_D^{(fund)}}
 \sigma_2^{(fund)}, C_D^{(2)} =\frac13 (\mu^2+\mu\nu+\nu^2+3\mu+3\nu).\label{5}\ee
 However for larger $n,n\geq 4$ the Casimir scaling is violated:
 \be \sigma_n^{(D)}= a_1C_D^{(2)} +a_2 (C^{(2)}_D)^2
 +a_3C^{(3)}_D+...\label{6}\ee
 It is remarkable that perturbative interaction of static quarks
 $V^{(D)}(r)$ satisfies the Casimir scaling to the order $O(g^6)$
 considered so far \cite{5}, so the total potential $V^{(D)}(r) =
 V^{(D)}_{pert} (r) + \sigma^{(D)} r+const$ is also Casimir
 scaling, if  GSA works well.

 This picture was tested recently on the lattice \cite{11} and
 confirmed the Casimir scaling with the accuracy around 1\% in the
 range $0.1\leq r\leq 1.1. $ fm. The full theoretical understanding
 of this fundamental fact is still lacking, both for the
 perturbative
 part and for the string tension. On the  pedestrian level the
 Casimir scaling and the quadratic (Gaussian) correlator
 dominance implies that  the vacuum  is highly stochastic and any
 quasiclassical objects, like instantons, are strongly suppressed
 in the real QCD vacuum. The vacuum consists of small white
 dipoles of the size $T_g$ made of neighboring field strength
 operators.  The smallness of $T_g$ might be an explanation for
 the  Gaussian dominance since higher correlator terms in $\sigma$
 are proportional to $(FT_g^2)^n (T_g^2)^{-1}$, where $F$ is the
 estimate of the average nonperturbative vacuum field, $F\sim 500
 $ (MeV)$^2$.

 Lattice calculations of FC have been done repeatedly during last
 decades, using cooling technic \cite{12} and with less
 accuracy without cooling \cite{13}. (Recently another  approach
 based on the  so-called gluelump states  was exploited on the
 lattice \cite{14} and analytically \cite{15}, which has a direct
 connection to FC).

The basic result of \cite{12} is  that FC  consists of
perturbative part $O(1/x^4)$ at  small distances and
nonperturbative part $O(\exp (-x/T_g))$ at larger distances  with
$T_g$ in the range $T_g = 0.2$ fm (quenched vacuum) and $T_g=0.3$
fm ( 2 flavours).

Calculations in \cite{13} and \cite{14,15}, as well as sum rule
estimates \cite{16} yield a smaller value, $T_g\approx 0.13$ fm to
0.17 fm. This enables us in what follows to take the limit $T_g\to
0$ keeping $\sigma = const \approx 0.18 $ GeV$^2$.

\section{Relativistic dynamics of confinement}

The Gaussian dominance and the small $T_g$ limit greatly simplify
the  relativistic dynamics of hadrons.

 As it was shown  both   for heavy quarks and  for
 light quarks (see \cite{6} for a review), in the limit of small
 $T_g$ one has local relativistic dynamics which can be described
 by a local Hamiltonian.
 Before entering into the
 details of the method, it is useful to list a number of problems,
 present in all dynamical calculations of hadrons, which have been
 easily solved in our Hamiltonian method:

 1) the problem  of the constituent quark and gluon mass

 2) the problem of Regge slope ($\alpha'=\frac{1}{8\sigma}$ for
 most calculations while in nature one has
 $\alpha'=\frac{1}{2\pi\sigma}$)

 3) the problem of Regge intercept (an arbitrary constant appears
 in the Hamiltonian in most calculations to shift the masses down
 to the experimental values).

 Finally, previous calculations are based on model assumptions and
 are not derived, in contrast to our  Hamiltonian directly from
 the QCD Lagrangian. This results in a large number of fitting
 parameters, whereas in our case we have none.

 The basic scheme of the Hamiltonian derivation consists of three
 steps:

 i) Definition of  initial and final states.

 ii) Exact representation of the gauge-invariant Green's function
 as a path integral over Wilson loops.

 iii) Derivation of the Hamiltonian from the  Green's function in
 the Gaussian approximation and the small $T_g$ limit.

We start with gauge-invariant asymptotic states, which are built
with the use of parallel transporters $\Phi(x,y)$ for nonlocal
states. This is done in the same way as on the lattice (but
without fuzzing links etc.) Thus for the meson one has
\be
\Psi_M(x,y) =\bar q(x) \Gamma\Phi (x,y) q(y),~~
\Gamma=(1,\gamma_\mu,...) \otimes 1,D_\nu, D_\mu D_\nu ...
\label{7}\ee Gluons can be created gauge-covariantly by
$F_{\mu\nu}$ and covariant derivatives e.g. $(D)^n F_{\mu\nu}$. At
this  point it is advantageous to separate perturbative (valence)
$a_\mu$ and nonperturbative  (background) gluonic field $B_\mu$,
\cite{17}
\be
A_\mu = B_\mu + a_\mu \label{8}\ee
 which are integrated independently in the partition function due
 to the 'tHooft identity \cite{17}
\be Z=\frac{1}{N} \int DA e^{-S(A)} =\frac{1}{N'} \int DB \int Da
 e^{-S(B+a)}.\label{9}
 \ee
 Assigning to $a_\mu$  homogeneous gauge transformations, one can
 create valence  gluonic states with the operators $a_\mu, (D)^n
 a_\mu$ etc.  where $D=\partial_\mu - ig B_\mu$. In this way
 glueball \cite{18}, hybrid \cite{6,19} and gluelump \cite{15} states
 have been introduced in good agreement with  lattice data.

 As the next step one introduces FFSR for the Green's function
 \cite{2}, for a review and earlier refs. see \cite{20}. E.g. for the meson it is
\be
G_{q\bar q}^\Gamma(x,y) = \lan G_q (,y) \Gamma G_{\bar q} ( x,y)
\Gamma- G_q(x,x) \Gamma G_{\bar q} (,y) \Gamma\ran_A\label{10}\ee
where FFSR for the (anti) quark Green's function is
\be
G_q(x,y) = (m-\hat D) \int^\infty_0 ds (Dz)_{xy} e^{-K}
\Phi_\sigma (x,y), K= \frac14 \int^\infty_0 \left
(\frac{dz_\mu}{d\tau}\right)^2 d\tau,\label{11} \ee

\be
\Phi_\sigma (x,y) = P_A\exp (ig \int^x_y A_\mu d z_\mu) P_F \exp
(g\int^s_0 d\tau \sigma_{\mu\nu} F_{\mu\nu})\label{12} \ee and
$(Dz)_{xy}$ implies the path integral from $y$ to $x$. Here $P_A,
P_F$ are ordering operators.
 A similar representation exists for the valence gluon
Green's function, with 3 changes; factor $(m-\hat D)$ in front in
(\ref{11}) is missing; $A_\mu, F_{\mu\nu}$ are in the adjoint
representation and $g\sigma_{\mu\nu} F_{\mu\nu}$ in (\ref{12})
should be replaced by $2g F_{\sigma\rho}$.

Insertion of (\ref{11}) into (\ref{10}) yields the main result to
be used below; all gluonic field is present in the gauge-invariant
form of the closed Wilson loop (modulo factor $(m-\hat D)$ and
$\sigma F$ insertions which are treated separately). Using for the
latter the area law, found on the lattice or analytically in the
Gaussian approximation, one has symbolically (neglecting spin
terms)
\be
G^\Gamma_{q\bar q} (x,y) \sim \int^\infty_0 ds \int^\infty_0 d\bar
s (Dz)_{xy} (D\bar z)_{xy} e^{-K-\bar K} e^{-\sigma
S_{min}(C)}.\label{13}\ee Here $S_{min} (C)$ is the minimal area
inside the contour $C$ made of the variable paths $C_z, C_{\bar
z}$ integrated in (\ref{13}). As the next step one goes over to
the Hamiltonian $H$ to escape from the path integrals, using the
standard definition
\be
G_{q\bar q}^\Gamma (x,y) =\lan x|e^{-HT}|y\ran.\label{14}\ee Here
$T$ is the evolution parameter associated with the chosen
hypersurface, i.e. for the c.m. Hamiltonian it is the c.m. time
coordinate. To proceed one should tackle the proper time $s$ (or
$\tau$) in (\ref{11}), (\ref{13}), and here enters a new
fundamental quantity which connects proper and actual (Euclidean)
time $t$,
\be
2\mu (t) = \frac{dt}{d\tau},~~ t\equiv z_4(\tau)\label{15}\ee so
that the integrals in (\ref{11}), (\ref{13}) can be rewritten as
(see \cite{20,21} for details)
\be
\int^\infty_0 ds (D^4z)_{xy}... = const \int D\mu (t)
(D^3z)_{xy}.\label{16} \ee Here $\mu(t)$ enters as an einbein
variable \cite{22},  physically the stationary point $\mu_0$ of
$\mu(t)$ will play the role of the constituent mass. Another
einbein variable $\nu(\beta)$ enters from the representation of
the string action in (\ref{13}) and has the physical meaning of
the string energy density along the string coordinate $\beta,
0\leq \beta \leq 1. $ The equal  quark mass $m_1=m_2=m$
Hamiltonian was obtained in \cite{21} and has the form

$$ H=\frac{p^2_r+m^2}{\mu(\tau)}+ \mu(\tau)+\frac{\hat L^2/r^2}
{\mu+2\int^1_0(\beta-\frac{1}{2})^2\nu(\beta) d\beta}+ $$ \be
+\frac{\sigma^2 r^2}{2}\int^1_0\frac{d\beta}{\nu(\beta)}+
\int^1_0\frac{\nu(\beta)}{2}d\beta, \label{17} \ee where
$p^2_r=(\vep\ver)^2/r^2$, and $L$ is the angular momentum, $\hat
L=(\ver\times \vep)$.

The physical meaning of the terms $\mu(t)$ and $\nu(\beta)$ can be
understood when one finds their extremal values. E.g. when
$\sigma=0$ and $L=0$, one finds from (\ref{17}) \be
H_0=2\sqrt{\vep^2+m^2},~~\mu_0=\sqrt{\vep^2+m^2} \label{18} \ee so
that $\mu_0$ is the energy of the quark. Similarly in the limiting
case $L\to \infty$ the extremum over $\nu(\beta)$ yields: \be
\nu_0(\beta)=\frac{\sigma r}{\sqrt{1-4y^2(\beta-\frac{1}{2})^2}},
~~M^2(L)=2\pi\sigma\sqrt{L(L+1)} \label{19} \ee so that $\nu_0$ is
the energy density along the string with the $\beta$ playing the
role of the coordinate along the string.

A similar form has the two-gluon (glueball) Hamiltonian, which
obtaines from (\ref{17}) putting $m\equiv 0$ and $\sigma \to
\sigma_{adj} =\frac94 \sigma$.

The generalization to more quarks and unequal masses and valence
gluons is straight-forward, one can find the corresponding
examples in \cite{6}.

\section{ Solution of the problems 1)-3)}

1) To find the stationary values of $\mu =\mu_0$ one can use an
approximate relation $\lan \frac{\partial H} {\partial \mu}
|_{\mu=\mu_0}\ran \approx \frac{\partial E(\mu)}{\partial \mu}
|_{\mu=\mu_0} =0$ which has a 5\% accuracy \cite{6}, and the
resulting  $\mu_0$ which can be considered as a   constituent
mass, is expressed entirely through $\sigma$. Taking $\sigma=0.18$
GeV$^2$ one obtains  for constituent quark mass from  the lowest
meson state $\mu_0^q$ (meson)=0.35 GeV and for the constituent
gluon mass from lowest  glueball state, $\mu_0^g$
(glueball)$=\frac32 \mu_0^q=0.52$ GeV. The proof that $\mu_0$ is
indeed constituent mass is obtained from the calculation of baryon
magnetic  moments which are expressed entirely through $\mu_0$ in
\cite{23}, and agree with experiments within 10\%.

2) The Regge-slope problem is actually  solved by (\ref{19}),
which yield $\alpha'=\frac{1}{2\pi\sigma}$ at large $L$ due to the
third term on the r.h.s. of (\ref{17}), containing the string
moment of inertia in the  denominator. This is just the term,
which is absent in the usual relativistic quark model (RQM)
predicting $\alpha'=\frac{1}{8\sigma}$. For not large $L$ the
Regge slope (and  the physical meson masses) have been calculated
in \cite{7,8} in good agreement with experimental meson masses and
$\alpha'_{\exp}$.

3) the Regge-intercept problem is connected to the fact that
actually the hadron mass is not given by the eigenvalue $M_0$ of
$H$ (\ref{17}), but rather by the renormalized  value
$M=M_0-\Delta M$. In all RQM calculations $\Delta M$ is considered
as a phenomenological parameter and is large, $\Delta M\sim 0.7$
GeV for mesons, whereas its physical origin was obscure.

In \cite{24} it was found that $\Delta M$ appears due to the
nonperturbative quark mass renormalozation, namely the selfenergy
due to the term $\sigma_{\mu\nu} F_{\mu\nu}$ in the second order.

This mechanism gives $\Delta M=\frac{4\sigma}{\pi}\sum^n_{i=1}
\frac{1}{2\mu_i} \eta_i$, where $\mu_i$ is the constituent mass of
the $i-$yh quark, the sum is over all quarks in the hadron and
$\eta_i(m_i)$ depends only on the current quark mass $m_i$ and
$\eta_i(0)=1$ for light quarks. This expression yields correct
values for $\Delta M$ for both mesons and baryons
\cite{7}-\cite{9}.

It is fascinating that i) $\mu_i$, entering the denominator of
$\Delta M$, help to keep intact the Regge-slope in $M=M_0-\Delta
M$. ii) the quark mass renormalization $\delta
m^2_q=-\frac{4\sigma}{\pi} $ is twice as large as in the 1+1 QCD.
Thus also the third problem -- the absolute values of hadron
masses -- is  simply solved within the method. The overall
agreement of calculated masses of light mesons in \cite{7,8},
heavy quarkonia in \cite{25}, baryons in \cite{9}, glueballs in
\cite{18}, hybrids in \cite{19} with experimental and/or lattice
data is better than 10\% and is surprising since the method
exploits only two inputs $\sigma$ and $\alpha_s$ (or
$\Lambda_{QCD}$) (in addition to current quark masses renormalized
at around 1 GeV).

Here we have touched only upon two basic topics in the framework
of the fast developing FCM, more details and applications can be
found in the review \cite{3} and cited literature.

The author is grateful to the Organizing Committee of Quarks -
2002 for kind hospitality and a nice scientific atmosphere at the
Conference. The financial support of the RFBR grant 00-02-17836 is
gratefully acknowledged.

 \end{document}